\documentclass[prl,twocolumn,preprintnumbers,amsmath,amssymb,superscriptaddress]{revtex4-1}

\usepackage{graphicx}
\usepackage{bm}
\usepackage{color}
\usepackage{textcase, natbib, url, ulem}
\usepackage[colorlinks=true, linkcolor=blue, urlcolor=blue, citecolor=blue]{hyperref}
\usepackage{amsmath,amssymb}
\usepackage{mathtools}

\DeclareMathOperator{\sgn}{sgn}

\begin{document}

\title{Transmission Phase in the Kondo Regime Revealed in a Two-path Interferometer}

\author{S. Takada}
\affiliation{Department of Applied Physics, University of Tokyo, Bunkyo-ku, Tokyo, 113-8656, Japan}
\author{C. B\"{a}uerle}
\affiliation{Universit\'{e} Grenoble Alpes, Institut NEEL, F-38042 Grenoble, France}
\affiliation{CNRS, Institut NEEL, F-38042 Grenoble, France}
\author{M. Yamamoto}
\affiliation{Department of Applied Physics, University of Tokyo, Bunkyo-ku, Tokyo, 113-8656, Japan}
\affiliation{PRESTO, JST, Kawaguchi-shi, Saitama 331-0012, Japan}
\author{K. Watanabe}
\affiliation{Department of Applied Physics, University of Tokyo, Bunkyo-ku, Tokyo, 113-8656, Japan}
\author{S. Hermelin}
\affiliation{Universit\'{e} Grenoble Alpes, Institut NEEL, F-38042 Grenoble, France}
\affiliation{CNRS, Institut NEEL, F-38042 Grenoble, France}
\author{T. Meunier}
\affiliation{Universit\'{e} Grenoble Alpes, Institut NEEL, F-38042 Grenoble, France}
\affiliation{CNRS, Institut NEEL, F-38042 Grenoble, France}
\author{A. Alex}
\affiliation{Physics Department, Arnold Sommerfeld Center for Theoretical Physics, and Center for NanoScience, Ludwig-Maximilians-Universit\"{a}t, Theresienstra\ss e 37, 80333 M\"{u}nchen, Germany}
\author{A. Weichselbaum}
\affiliation{Physics Department, Arnold Sommerfeld Center for Theoretical Physics, and Center for NanoScience, Ludwig-Maximilians-Universit\"{a}t, Theresienstra\ss e 37, 80333 M\"{u}nchen, Germany}
\author{J. von Delft}
\affiliation{Physics Department, Arnold Sommerfeld Center for Theoretical Physics, and Center for NanoScience, Ludwig-Maximilians-Universit\"{a}t, Theresienstra\ss e 37, 80333 M\"{u}nchen, Germany}
\author{A. Ludwig}
\affiliation{Lehrstuhl f\"{u}r Angewandte Festk\"{o}rperphysik, Ruhr-Universit\"{a}t Bochum, Universit\"{a}tsstra\ss e 150, 44780 Bochum, Germany}
\author{A. D. Wieck}
\affiliation{Lehrstuhl f\"{u}r Angewandte Festk\"{o}rperphysik, Ruhr-Universit\"{a}t Bochum, Universit\"{a}tsstra\ss e 150, 44780 Bochum, Germany}
\author{S. Tarucha}
\affiliation{Department of Applied Physics, University of Tokyo, Bunkyo-ku, Tokyo, 113-8656, Japan}
\affiliation{RIKEN Center for Emergent  Matter Science (CEMS), 2-1 Hirosawa, Wako-shi, Saitama 31-0198, Japan}

\date{\today}

\begin{abstract}
We report on the direct observation of the transmission phase shift through a Kondo correlated quantum dot by employing a new type of two-path interferometer.
We observed a clear $\pi/2$-phase shift, which persists up to the Kondo temperature $T_K$.
Above this temperature, the phase shifts by more than $\pi/2$ at each Coulomb peak, approaching the behavior observed for the standard Coulomb blockade regime.
These observations are in remarkable agreement with two-level numerical renormalization group calculations.
The unique combination of experimental and theoretical results presented here fully elucidates the phase evolution in the Kondo regime.
\end{abstract}



\maketitle


The Kondo effect, an archetype of many-body correlations, arises from the interaction between a localized spin and surrounding conduction electrons \cite{JKondo1964}.
It is characterized by a many-body singlet ground state, often referred to as the Kondo cloud.
It was first observed in metals with a small inclusion of magnetic impurities and manifests itself by a logarithmic increase of the electrical resistance at low temperatures.
Recently, the advance in nanotechnology has made it possible to study the interaction of a single impurity spin in contact with an electron reservoir, in particular in semiconductor quantum dots (QDs) \cite{Nature391_156, Science289_2105}.
These developments have also allowed access to the phase shift across a Kondo impurity, a central ingredient of Nozi\`{e}res' celebrated Fermi-liquid theory for the low-energy fixed point of the Kondo effect \cite{Nozieres:xy}: when an electron of sufficiently low energy is incident on the impurity in the screened singlet state, it scatters coherently off the latter, acquiring a $\pi/2$-phase shift, but with zero probability for a spin flip.
This $\pi/2$-phase shift is one of the hallmarks of the Kondo effect.
Though it cannot be measured directly in conventional Kondo systems, it has been suggested \cite{Gerland2000} to be possible for a semiconductor QD placed in one arm of an Aharonov-Bohm (AB) ring, where the transmission phase through the dot can be extracted from the AB oscillations of the conductance as a function of the magnetic flux through the ring \cite{Yacoby1995, Schuster1997}.
For such geometries it has been predicted that the $\pi/2$-phase shift should persist, perhaps surprisingly, even up to temperatures as high as the Kondo temperature, $T_K$ \cite{Silvestrov2003, Hecht2009}.

Such Kondo phase measurements have indeed been pursued in a number of remarkable pioneering experiments \cite{Ji2000, Ji2002, Zaffalon2008, Sato2005, Katsumoto2006}.
However, unexpected results have been obtained for the phase shift in the Kondo regime ($T\lesssim T_K$): for instance, Ref.~\onlinecite{Ji2000} reported a $\pi$ plateau in the Kondo valley and the total phase change through the spin-degenerate pair was of about 1.5$\pi$, while Ref.~\onlinecite{Ji2002} reported an almost linear phase shift over a range of $1.5\pi$ when the Fermi energy was scanned through a spin degenerate level.
Explanations of these surprising features are still lacking. 
The $\pi/2$ plateau in the Kondo valley has only been reported in Ref.~\onlinecite{Zaffalon2008}. In that measurement, however, the temperature was so large ($T/T_K \gtrsim 30$) that Kondo correlations are strongly suppressed.
In such conditions one does not expect to observe the $\pi/2$-Kondo phase shift, which is predicted to occur only for $T < T_K$ \cite{Gerland2000}.
Another puzzling feature of this experiment is that the $\pi/2$-phase shift extends over the entire Kondo valley.
According to Kondo theory \cite{Gerland2000,Hecht2009}, one would actually expect a strong phase dependence within the Kondo valley for the experimental conditions of Ref.~\onlinecite{Zaffalon2008} (see Ref.~\onlinecite{suppl-mat}).

In this Letter, we report on measurements of the transmission phase shift through a Kondo QD using an original two-path interferometer.
We confirm the $\pi$-phase shift across a Coulomb peak (CP) without Kondo correlation as expected from a Breit-Wigner type resonance.
In the Kondo regime, we observe a clear $\pi/2$-phase shift at $T \lesssim T_K$.
Increasing $T$ above $T_K$, the phase shift starts to deviate from the $\pi/2$ behavior and approaches the one observed for the standard Coulomb blockade regime.
Via complementary two-level numerical renormalization group calculations \cite{Wilson1975,Weichselbaum2007,Bulla2008,Hecht2009,Weichselbaum2012} we obtain a detailed and consistent understanding of the phase behavior in the Kondo regime.

The measurement of the true transmission phase through a QD in an AB setup is not at all trivial. 
Multiple path contributions by electrons that encircle the AB loop multiple times alter the phase of the observed AB oscillation and a reliable extraction of the \textit{true} transmission phase through a QD is impossible.
The most conspicuous example is the so-called phase rigidity \cite{Buttiker1988, Yacoby1996} in a two-terminal setup.
It has been shown that employing multi-terminal devices can lift the phase rigidity \cite{Ji2000, Ji2002, Zaffalon2008, Schuster1997, Sigrist2004}.
However, it is far from obvious that in such a device multiple path contributions are sufficiently suppressed. In contrast, with our original two-path interferometer we are able to suppress such multiple path contributions.

\begin{figure}[htbp]
	\includegraphics[width=0.5\textwidth]{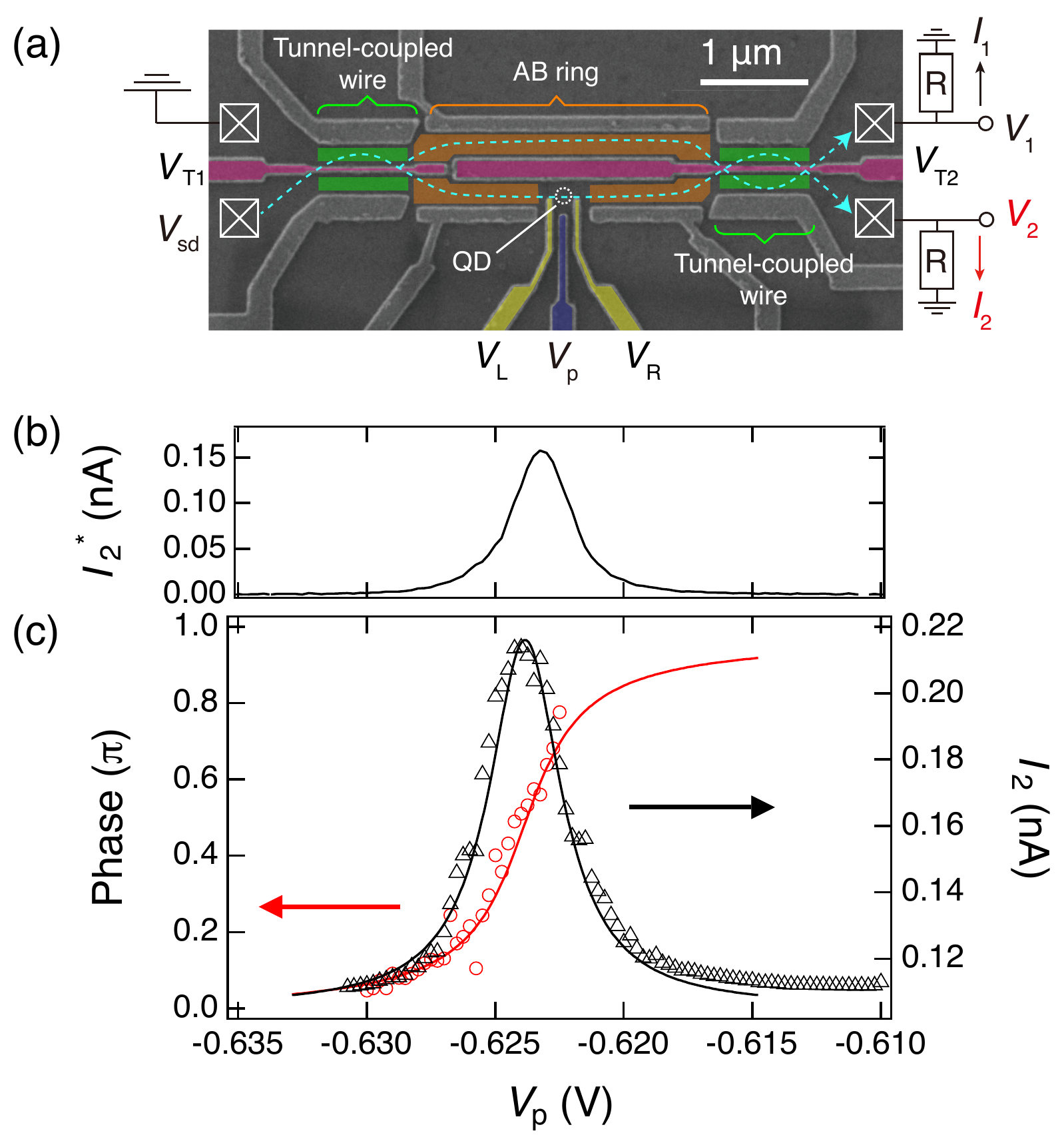}
	\caption{\label{fig:sample}
(a) Scanning electron micrograph of the employed device and measurement setup. The AB ring is connected to a tunnel-coupled wire on both sides. A QD is embedded into the lower path of the AB ring. The dashed lines indicate the two paths. The current flowing through the upper (lower) contact is obtained from $I_{\rm 1(2)} = V_{\rm 1(2)}/R$ ($R=10 \ {\rm k\Omega}$).
(b) Coulomb peak of the QD in the Coulomb blockade regime, where $I_{\rm 2}^*$ corresponds to the current passing only through the QD by completely depleting the left tunnel gate $V_{T1}$. The energy level of the QD is scanned across the Fermi energy by changing the plunger gate voltage $V_p$. A charging energy $U \sim 1.8{\rm ~meV}$, a tunnel-coupling energy $\Gamma \sim 40{\rm ~\mu eV}$, and a single level spacing $\delta \sim 230{\rm ~\mu eV}$ were estimated by mapping out the Coulomb diamond (not shown here).
(c) Phase evolution across the Coulomb peak extracted from the antiphase oscillations is plotted on the left axis (red circles). The current collected in $I_{\rm 2}$ is plotted on the right axis (black triangles), where $I_{\rm 2}$ is averaged over one oscillation period of magnetic field. For gate voltages more positive than $-0.623$V, charging noise did not allow us to follow the oscillations and reliable data of the transmission phase could not be obtained. Black and red solid lines show Lorentzian fitting of $I_{\rm 2}$ and the transmission phase expected from Friedel's sum rule (see text), respectively}.
\end{figure}
\begin{figure}[htbp]
	\includegraphics[width=0.45\textwidth]{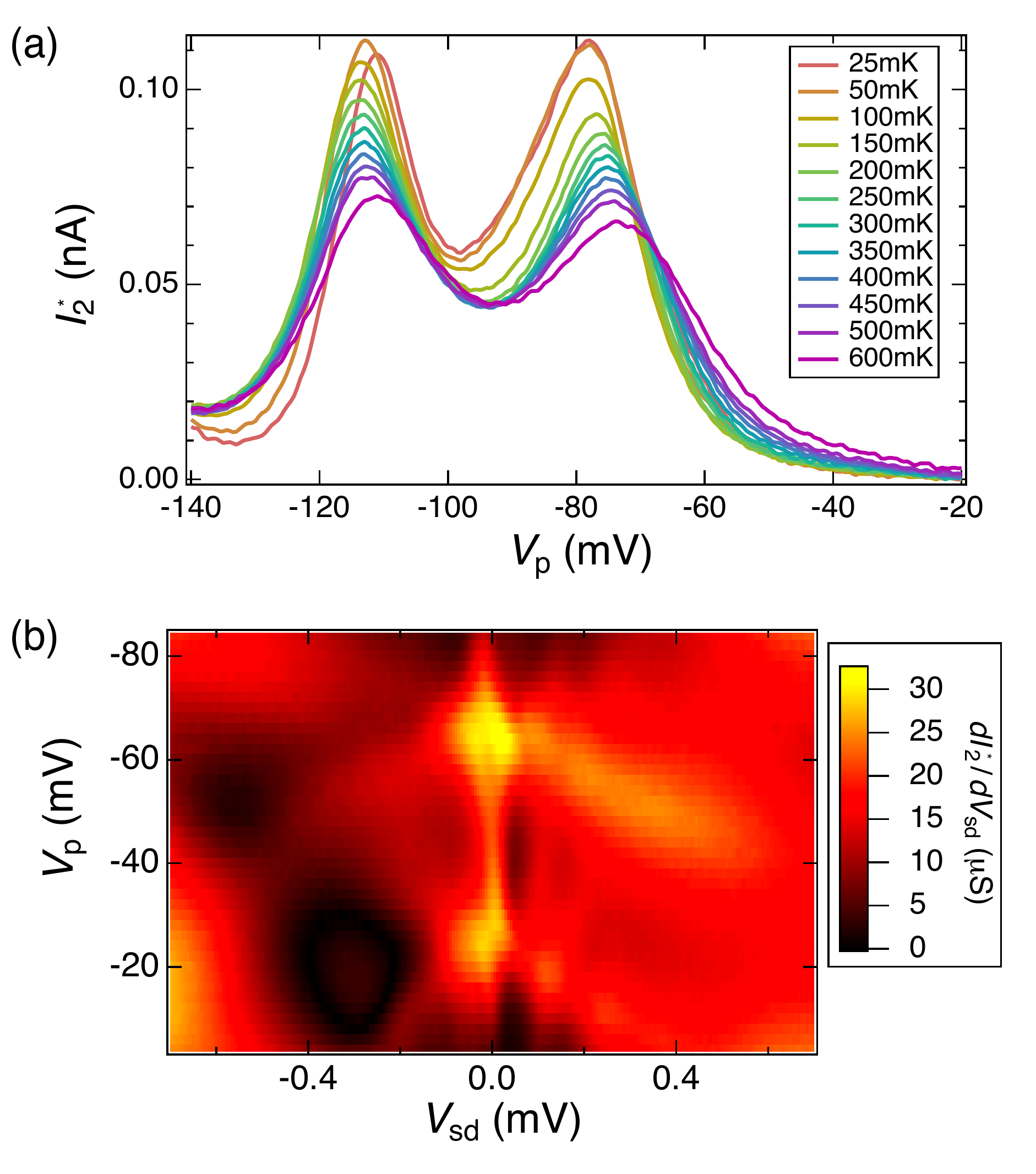}
	\caption{\label{fig:Kondo}(a) Temperature dependence of two Coulomb peaks which show Kondo correlations, observed in $I_{\rm 2}^*$ plotted as a function of $V_p$, where the left tunnel gate $V_{T1}$ was depleted. Here $V_{\rm sd} \sim 3.5 \ {\rm \mu}{\rm V}$ was applied to measure the current.
(b) Coulomb diamond for the CPs shown in (a) in the Kondo regime ($T \lesssim T_K$). The gate voltage configuration in (b) differs from that in (a) by a $V_p$ shift of $\sim 50$ mV, because (b) was measured using slightly wider one-dimensional channels for the leads, to get rid of some additional $V_p$ independent structures in the Coulomb diamond coming from the narrow one-dimensional channels \cite{Kristensen2000}, which disappeared upon slightly widening the channels.}
\end{figure}
\begin{figure*}[htbp]
	\includegraphics[width=0.9\textwidth]{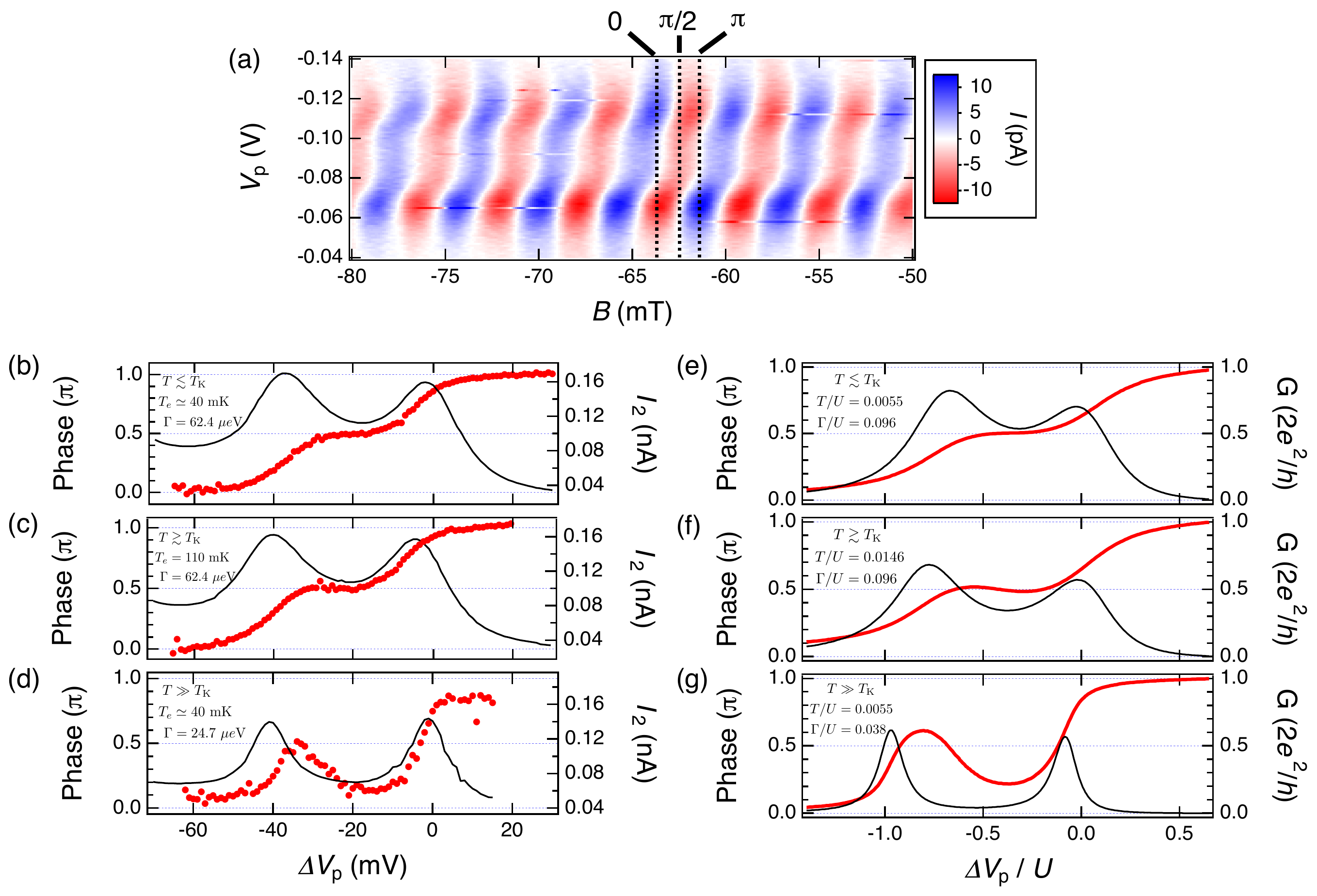}
\caption{\label{fig:KondoPhase}
(a) AB oscillation amplitude as a function of magnetic field, obtained for the same gate configurations as Fig.~\ref{fig:Kondo}(a) except for $V_{T1}$ by scanning the plunger gate voltage $V_p$ across two successive Coulomb peaks, showing Kondo correlations in the valley between them ($T \lesssim T_K$). The AB oscillation amplitude is obtained from $I = I_{\rm 1} - I_{\rm 2}$ by subtracting a smoothed background. Dotted lines emphasize the phase shift.
(b), (c), (d) The transmission phase (red circles, left axis), determined by a complex FFT of the antiphase oscillations, together with $I_{\rm 2}$ (black line, right axis), plotted as a function of the change $\Delta V_p$ in plunger gate voltage with respect to its value at the center of the right CP, for three different regimes, (b): $T \lesssim T_K$, (c): $T \gtrsim T_K$ and (d): $T \gg T_K$. $I_{\rm 2}$ is averaged over one oscillation period of magnetic field. Parameters shown in the figures are estimates, extracted from fitting NRG results to the measured phase curves. (e), (f), (g) Transmission phase (red line) and conductance (black line) calculated by a two-level Anderson impurity model with $s=+$ and the lower level being the Kondo level, for an interaction energy $U=650{\rm ~\mu eV}$ and a level spacing of $\delta = 0.5 U$ (for details, see Ref.~\onlinecite{suppl-mat}). Fitting parameters used for the calculations are shown in the figures. Current $I_{\rm 2}$ of the experimental data (b), (c), (d) contains a linear background from the upper arm of the interferometer whereas the conductance $G$ in the theoretical calculations (e), (f), (g) is the bare conductance across a QD.
}
\end{figure*}

The device was fabricated in a two-dimensional electron gas [2DEG, $n = 3.21 \times 10^{11}{\rm ~cm^{-2}}$, $\mu = 8.6 \times 10^5 {\rm ~cm^2 / Vs}$ at $4.2$ K; see Fig.~\ref{fig:sample}(a)] formed at an AlGaAs/GaAs heterointerface with standard surface Schottky gate technique.
It consists of a novel AB interferometer, where the AB ring is attached to two tunnel-coupled wires,  and acts as a \textit{true} two-path interferometer when the tunnel-coupled wires are set to work as half beam splitters by carefully tuning the tunnel gate voltages $V_{T1}$ and $V_{T2}$ \cite{Yamamoto:2012fk}.
This does not degrade the current oscillation signal (maximum visibility $\sim 15 \%$) in contrast with multiterminal devices employed in previous works, where most of the coherent electrons are lost in the base contacts.
The unique advantage of this device is that the true transmission phase can be determined with very high accuracy since contributions from the multipath interferences can be fully eliminated by simply tuning the gate voltages such that the two output currents $I_{\rm 1}$ and $I_{\rm 2}$ oscillate with opposite phase as a function of the magnetic field \cite{Yamamoto:2012fk}.
In this case the total current $(I_{\rm 1}+I_{\rm 2})$ through the interferometer shows no magnetic field dependence, a signature that loop trajectories are suppressed, leaving only the two-path interference.
The phase extracted from the antiphase oscillations of $I_{\rm 1}$ and $I_{\rm 2}$ is always reproducible and consistent with theoretical expectations.
In contrast, when $I_{\rm 1}$ and $I_{\rm 2}$ are not antiphase due to the existence of multiple loops, the phase of the current oscillation becomes different from the reproducible and consistent one even if it smoothly shifts with the voltages on one of the gates forming the AB ring, suggesting the possibility that previous experiments \cite{Ji2000, Ji2002, Zaffalon2008} might still have suffered from multiple path contributions \cite{Entin2002, Aharony2005, Simon2005, Aharony2006}.
All the measurements were done in a dilution refrigerator at a base temperature of approximately 15 mK, except for Fig.~\ref{fig:Kondo}(a) and Fig.~\ref{fig:KondoPhase}(c).
For the transport measurements an ac bias ($3 \sim 20 {\rm ~\mu V}, \ f = 23.3 {\rm ~Hz}$) was applied to the lower left contact and currents $I_{\rm 1}$ and $I_{\rm 2}$ were measured at the two output contacts using a standard lock-in technique.
This bias is sufficiently small that the current response is linear to deduce the differential conductance of the system studied here.

To characterize the sample, we first measured the transmission phase shift across a CP without Kondo correlation, where the temperature is much larger than $T_K$.
Transport only through the QD was achieved by depleting the left tunnel gate $V_{T1}$; Fig.~\ref{fig:sample}(b) shows the corresponding current $I_{\rm 2}^*$.
For the phase measurement, the left tunnel-coupled wire was then tuned to a half beam splitter.
For each value of $V_p$ the magnetic field dependencies of the two output currents $I_{\rm 1}$ and $I_{\rm 2}$ are recorded.
In order to extract the oscillating part, we subtract the smoothed background from the raw data \cite{Chandrasekhar1991} (see Ref.~\onlinecite{suppl-mat}).
The dependence of the transmission phase shift on $V_p$ is then obtained by performing a complex fast Fourier transform (FFT) of the oscillating part \cite{Heiblum_phase_2005, Chandrasekhar1991, Zaffalon2008}.
Other gate voltages are fixed during the measurement.

The extracted phase is shown in Fig.~\ref{fig:sample}(c).
The shape of the current $I_{\rm 2}$ mimics that of the CP $I_{\rm 2}^{*}$~, apart from an additional finite background current due to the current flowing through the upper path.
Here we show the nonoscillating part of $I_{\rm 2}$ averaged over one oscillation period of the magnetic field.
The change in the transmission phase $\Delta \varphi_{\rm dot}$ with changing plunger gate voltage is related to the corresponding change in QD charge (in units of the electron charge) $\Delta N$, via Friedel's sum rule,
$\Delta N= \Delta \varphi_{\rm dot}/\pi$ \cite{PhysRevB.52.R14360}.
The red solid line in Fig.~\ref{fig:sample}(c) shows the phase extracted from the averaged $I_{\rm 2}$ via Friedel's sum rule. 
First, $I_{\rm 2}$ is fitted by a Lorentzian function to obtain a continuous line shape of the CP (black solid line), which also gives the total area of the CP.
$\Delta N$ as a function of $V_p$ is then given by the integrated area under the fitted CP up to $V_p$ divided by the total area of the CP.
The measured phase shift shown by the red circles across the CP is in very good agreement with the phase extracted using Friedel's sum rule \cite{PhysRevB.52.R14360}.

We then focused on a pair of CPs and tuned them into the Kondo regime by $V_L$, $V_p$, $V_R$ [see Fig.~\ref{fig:sample}(a)].
We first characterized the QD using transport measurements only through the dot.
Panel (a) of Fig.~\ref{fig:Kondo} shows the temperature dependence of two CPs, where the Kondo effect appears in the valley between the two peaks around a plunger gate voltage of $-100$ mV.
The current $I_{\rm 2}^*$ in the valley increases as the temperature is lowered, a clear signature of the Kondo effect.
As we do not reach the unitary limit, $T_K$ cannot be determined with high precision from this temperature dependence of the conductance, but we find an upper bound of approximately $100\ {\rm mK}$ for the Kondo temperature at the valley center.
Figure~\ref{fig:Kondo}(b) shows a Coulomb diamond in this regime and a clear maximum (vertical yellow ridge) is observed around zero bias in the Coulomb blockade region, usually referred to as the zero bias anomaly.

The transmission phase across these two CPs was then measured by opening the left tunnel gate $V_{T1}$, and recording the AB oscillations for different plunger gate voltages $V_p$, as shown in Fig.~\ref{fig:KondoPhase}(a).
In this experiment, the energy scale of the magnetic field was lower than that of the temperature, so that spin was not resolved \cite{PhysRevB.86.115129}; however, we see clear signatures of the Kondo effect.
The visibility of our AB oscillations is sufficiently large to reveal, even in the raw data, a clear phase shift of approximately $\pi/2$ in the Kondo valley, as indicated by the dotted lines in Fig.~\ref{fig:KondoPhase}(a).
To reveal the phase shift across the CPs even more clearly, we performed a complex FFT as described above.
The obtained transmission phase is presented in Fig.~\ref{fig:KondoPhase}(b), together with $I_{\rm 2}$.
A total phase shift of $\pi$ when scanning through the two successive CPs is observed and a clear plateau appears at $\pi/2$ in the Kondo valley.
This result contrasts with the phase shift across a CP without Kondo correlation.
Although the conductance does not reach the unitary limit, the transmission phase shift in the Kondo valley already shows the $\pi/2$-phase shift, which is consistent with theoretical predictions that the $\pi/2$-phase shift survives for temperature as large as $T \sim T_K$ \cite{Silvestrov2003, Hecht2009}.

When increasing the temperature to approximately $110$ mK, the Kondo correlations are reduced ($T \gtrsim T_K$) and the behavior of the transmission phase is slightly altered.
The phase climbs slightly above $\pi/2$ before decreasing slightly, forming an $S$-shape [Fig.~\ref{fig:KondoPhase}(c)] in the Kondo valley.
To further reduce the Kondo correlations, we reduced the coupling energy $\Gamma$ by suitably tuning $V_L$, $V_p$, $V_R$, which lowers the Kondo temperature [Fig.~\ref{fig:KondoPhase}(d)], thus reaching a regime where $T \gg T_K$.
In this regime, the $S$-shaped structure of the phase evolution is more pronounced and becomes asymmetric with respect to the valley center and the value $\varphi_{\rm dot} = \pi/2$.
This suggests that by further reducing the Kondo correlations the phase behavior across each CP would approach that observed for CPs without Kondo correlation.
These results differ strikingly from the observation of Ref.~\onlinecite{Zaffalon2008}, where a $\pi/2$-phase shift has been reported for a regime where $T \gg T_K$ (see Ref.~\onlinecite{suppl-mat}).

To further corroborate our findings we performed numerical renormalization group (NRG) calculations of the conductance as well as of the transmission phase.
We found that even when the single level spacing $\delta$ is larger than the coupling energy $\Gamma$, the contribution from multiple levels in the dot plays an important role for the phase behavior at $T \gg T_K$. 
Taking into account only a single level results in a phase evolution, which is symmetric with respect to the center of the Kondo valley, contrary to our experimental findings [see Fig.~\ref{fig:KondoPhase}(d)].
In order to accurately reproduce the experimental phase data, we therefore employed a two-level Anderson model (see Ref.~\onlinecite{suppl-mat}). It is crucial to choose correctly the relative sign of the tunnel-coupling coefficients [i.e.,\ $s=\sgn(t^1_L t^1_R t^2_L t^2_R)$] for fitting the data of Fig.~\ref{fig:KondoPhase}(d), since this sign characterizes the influence of the nearby orbital level on the phase evolution.
Fitting the predicted phase evolution to the one observed experimentally (see Ref.~\onlinecite{suppl-mat}) allows us to estimate the QD parameters, such as $\Gamma$ and $\delta$, with good precision (note that these are not easily accessible only from the measured conductance data, due to renormalization of the CPs by the Kondo effect).
In addition, the slope of the phase in the Kondo valley in Fig.~\ref{fig:KondoPhase}(c) allows us to precisely evaluate the Kondo temperature at the valley center to $T_K \simeq 50$ mK for Figs. \ref{fig:KondoPhase}(b) and \ref{fig:KondoPhase}(c).
From the temperature evolution of the $\pi/2$ plateau, which is most prominent close to $T_K$, we are able to evaluate the actual electron temperature to $T_e \sim 40$ mK, consistent with previous measurements in the same electromagnetic environment \cite{PhysRevB.81.245306}.
The most important finding, however, is the fact that the $\pi/2$-phase shift persists up to a temperature of $T_K$ and then evolves into an $S$ shape at higher temperatures, which is extremely well captured by the NRG calculations. 

In summary, our measurements show that the transmission phase through a Kondo QD is $\pi/2$ up to temperatures of the order of $T_K$, in full agreement with theoretical expectations.
The key new ingredient underlying this result is the use of a novel AB interferometer design that can be fine-tuned to reliably exclude the contributions of paths traversing the interferometer multiple times.
This new design, in combination with detailed theoretical calculations for a two-level Anderson model, allowed us to obtain a detailed and consistent understanding of the behavior of the Kondo phase shift.

\begin{acknowledgements}
S. Takada acknowledges support from JSPS Research Fellowships for Young Scientists and French Government Scholarship for Scientific Disciplines. C.B. acknowledges financial support from the French National Agency (ANR) in the frame of its program BLANC “FLYELEC” Project No. anr-12-BS10-001 as well as from DRECI-CNRS/JSPS (PRC0677) International collaboration. M.Y. acknowledges financial support by Grant-in-Aid for Young Scientists A (No. 23684019) and Grant-in-Aid for Challenging Exploratory Research (No. 25610070). A.A., A.W., and J.v.D. acknowledge support from the DFR through SFB-631, SFB-TR12 and the Cluster of Excellence Nanosystems Initiative Munich, and from the Center for NanoScience. A.L. and A.D.W. acknowledge support from Mercur Pr-2013-0001, BMBF - Q.com-H  16KIS0109, and the DFH/UFA  CDFA-05-06. S.Tarucha acknowledges financial support by JSPS, Grant-in-Aid for Scientific Research S (No. 
26220710), MEXT KAKENHHI “Quantum Cybernetics", MEXT project for Developing Innovation Systems, and JST Strategic International Cooperative Program.
\end{acknowledgements}

%

\onecolumngrid
\newpage

\renewcommand{\thefigure}{{\bf S\arabic{figure}}}
\makeatother
\setcounter{figure}{0}
\setcounter{page}{1}
\thispagestyle{empty}

\begin{center}
\textbf{{\large Supplemental Material for Transmission Phase in the Kondo Regime Revealed in a Two-Path Interferometer}}\\
\bigskip
S. Takada,$^{\rm 1}$ C. B\"{a}uerle,$^{\rm 2, 3}$ M. Yamamoto,$^{\rm 1, 4}$ K. Watanabe,$^{\rm 1}$ S. Hermelin,$^{\rm 2, 3}$ T. Meunier,$^{\rm 2, 3}$\\ A. Alex,$^{\rm 5}$ A. Weichselbaum,$^{\rm 5}$ J. von Delft,$^{\rm 5}$ A. Ludwig,$^{\rm 6}$ A. D. Wieck,$^{\rm 6}$ and S. Tarucha$^{\rm 1, 7}$\\
$^{\rm 1}$\textit{Department of Applied Physics, University of Tokyo, Bunkyo-ku, Tokyo, 113-8656, Japan}\\
$^{\rm 2}$\textit{Universit\'{e} Grenoble Alpes, Institut NEEL, F-38042 Grenoble, France}\\
$^{\rm 3}$\textit{CNRS, Institut NEEL, F-38042 Grenoble, France}\\
$^{\rm 4}$\textit{PRESTO, JST, Kawaguchi-shi, Saitama 331-0012, Japan}\\
$^{\rm 5}$\textit{Physics Department, Arnold Sommerfeld Center for Theoretical Physics, and Center for NanoScience, Ludwig-Maximilians-Universit\"{a}t, Theresienstra\ss e 37, 80333 M\"{u}nchen, Germany}\\
$^{\rm 6}$\textit{Lehrstuhl f\"{u}r Angewandte Festk\"{o}rperphysik, Ruhr-Universit\"{a}t Bochum, Universit\"{a}tsstra\ss e 150, 44780 Bochum, Germany}\\
$^{\rm 7}$\textit{RIKEN Center for Emergent Matter Science (CEMS), 2-1 Hirosawa, Wako-shi, Saitama 31-0198, Japan}
\end{center}
\vspace{8pt}
\twocolumngrid

\section{\label{sec:s1}Two-path interference in an AB ring with tunnel-coupled wires}
\begin{figure}[htbp]
	\includegraphics[width=0.45\textwidth]{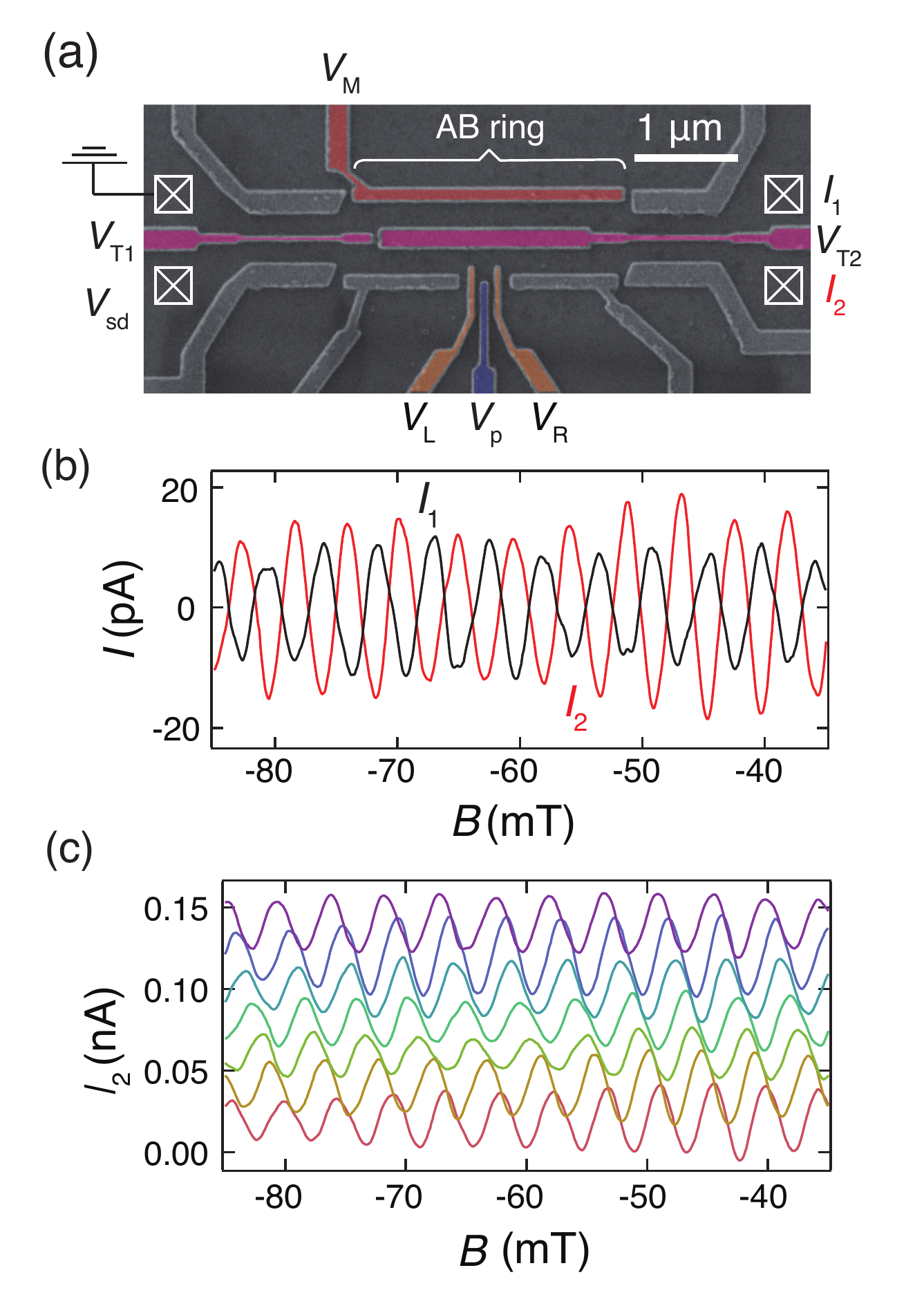}
	\caption{\label{fig:sampleb}(a) A scanning electron micrograph of the employed device. The AB ring is connected on both sides to a tunnel-coupled wire. A QD is embedded into the lower path of the AB ring. (b) Magnetic field dependence of the currents $I_{\rm 1}$ and $I_{\rm 2}$. The oscillating part of the currents are extracted by subtracting a smoothed background from the raw data. (c) Magnetic field dependence of the current $I_{\rm 2}$ for several values of the gate voltage $V_M$. $V_M$ is stepped by $5$ mV from $-0.24$ V to $-0.275$ V and each curve is shifted vertically for clarity.}
\end{figure}
For the measurement we utilized an original Aharonov-Bohm (AB) interferometer, where the AB ring is connected on both sides to a tunnel-coupled wire [Fig. \ref{fig:sampleb}(a)].
It has been shown that this device works as a two-path interferometer when the tunnel-coupled wires are set to work as half beam splitters by finely tuning the gate voltages $V_{T1}$ and $V_{T2}$ in a suitable magnetic field range \cite{Yamamoto:2012fkb}.
Assuming that transport is coherent and mediated by only a single transmitting channel, the output currents $I_{\rm 1}$ and $I_{\rm 2}$ are given, respectively,
\begin{eqnarray}
	I_{\rm 1(2)} = \frac{2e^2}{h} V_{\rm sd} \cdot \frac{[1\pm \cos \theta]}{2} \ ,	\label{eq:anti}
\end{eqnarray}
where $V_{\rm sd}$ is the bias applied to the lower left contact and $\theta$ is the phase difference accumulated between the upper and lower paths through the AB ring.
$\theta = \oint {\bm k} \cdot {\rm d}{\bm l} - \frac{e}{\hbar} BS + \varphi_{\rm dot}$ is composed of three terms.
The first term is the geometrical phase originating from the motion of electrons on a closed geometrical path around the ring and is obtained by integrating the wave vector of electrons ${\bm k}$ over the electron path.
The second term is the AB phase, where $B$ is the magnetic field applied perpendicularly to the surface and $S$ is the area enclosed by the two paths of the AB ring.
The third term is the transmission phase accumulated across the quantum dot (QD) embedded in the lower arm of the AB ring.

Figure \ref{fig:sampleb}(b) shows the oscillating part of $I_{\rm 1}$ and $I_{\rm 2}$ as a function of the magnetic field, or in other words, the modulation of the AB phase term of $\theta$.
$I_{\rm 1}$ and $I_{\rm 2}$ oscillate with opposite phase as expected from Eq.(\ref{eq:anti}).
In our experiment the visibility defined by the amplitude of the antiphase oscillation divided by the average current were typically $5 \sim 15 \%$.
Opposite phase indicated that the total transmission through the interferometer appearing as total current $I_{\rm 1} + I_{\rm 2}$ is almost independent for $\theta$.
The influences of multiple loops in the ring are therefore highly suppressed. 
When such antiphase oscillations are observed, we can see that $\theta$ is also modulated by the gate voltage $V_{M}$ [see Fig. \ref{fig:sampleb}(a)] via the geometrical phase.
These results are shown in Fig. \ref{fig:sampleb}(c), where the oscillation of $I_{\rm 2}$ plotted in different colors were measured for different gate voltages of $V_{M}$.
The phase of the oscillation smoothly shifts by about $2\pi$ with $V_{M}$.
Antiphase oscillation of two output currents and the smooth phase shift for $V_{M}$ indicate that the phase of the observed oscillation is not perturbed by the contribution from multipath interference but simply reflects $\theta$ accumulated through the AB ring in the \textit{true} two-path interference.
Measuring either $I_{\rm 1}$ or $I_{\rm 2}$ in a regime where they are antiphase allows for unambiguous measurement of the transmission phase through the quantum dot.

\section{\label{sec:subtraction}Analysis of phase shift}
\begin{figure}[htbp]
	\includegraphics[width=0.45\textwidth]{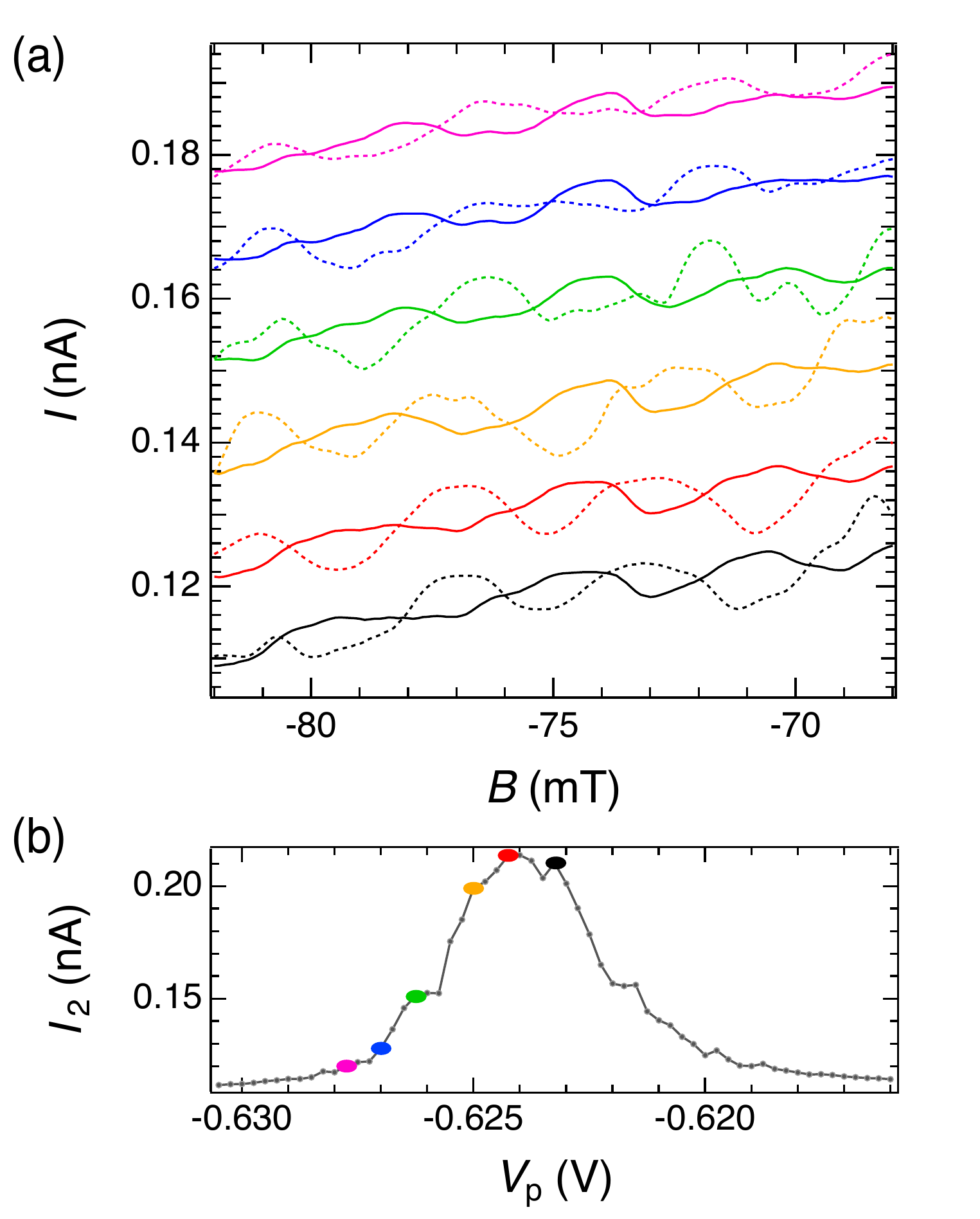}
	\caption{\label{fig:rawCoulomb}(a) Raw data of two output currents $I_{\rm 1}$ (solid lines) and $I_{\rm 2}$ (broken lines) as a function of the magnetic field at six values of $V_{p}$ of Fig. 1(c). Each data curve is shifted vertically for clarity. (b) The corresponding $I_{\rm 2}$ at each $V_{p}$ of (a) averaged over one oscillation period from $B=73$ mT to $-68.5$ mT. The colored dots correspond to the colored traces in (a).}
\end{figure}
\begin{figure}[htbp]
	\includegraphics[width=0.45\textwidth]{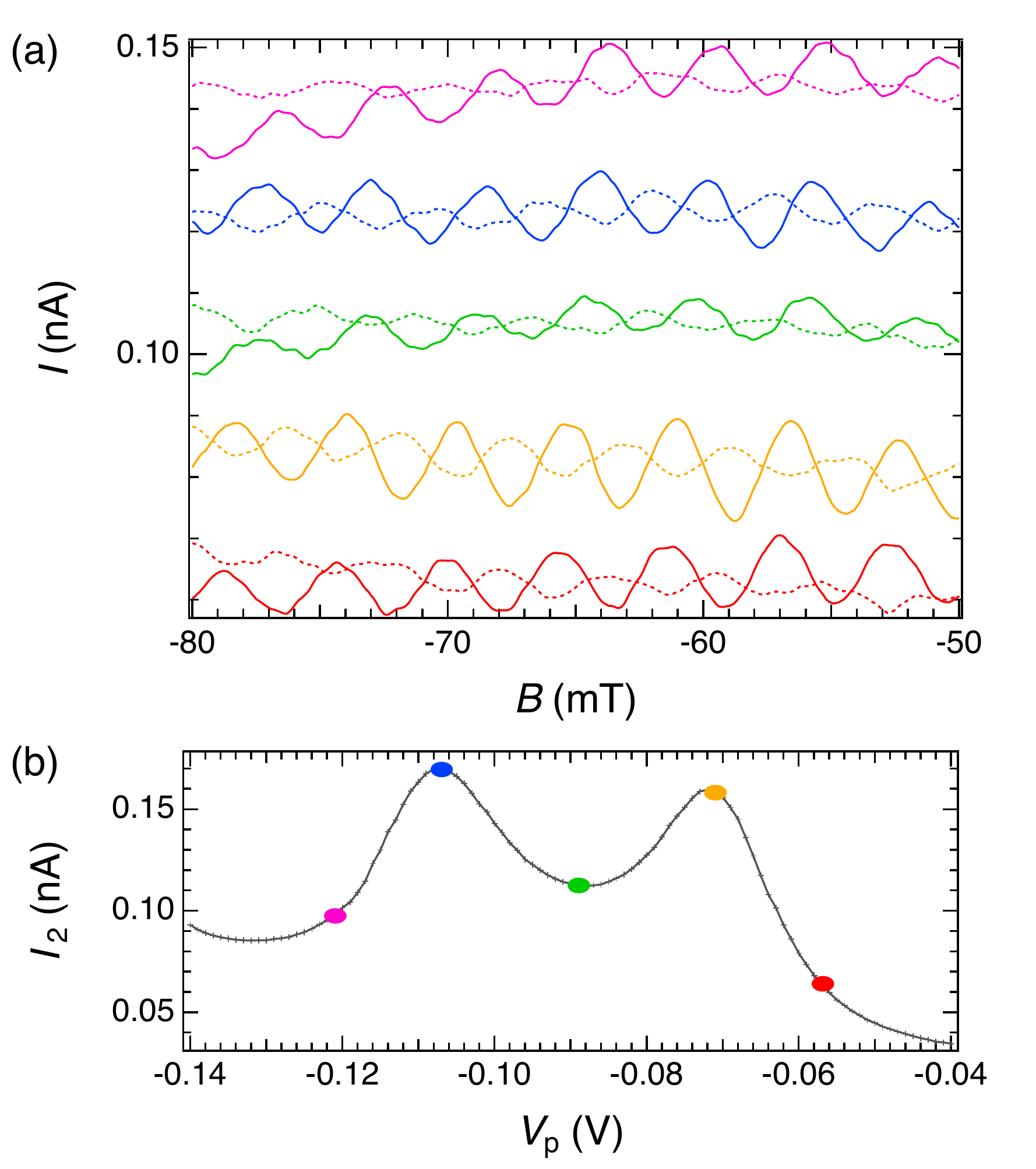}
	\caption{\label{fig:rawKondo}(a) Raw data of two output currents $I_{\rm 1}$ (solid lines) and $I_{\rm 2}$ (broken lines) as a function of the magnetic field at six values of $V_{p}$ of Fig. 3(a) and 3(b). Each data curve is shifted vertically for clarity. (b) The corresponding $I_{\rm 2}$ at each $V_{p}$ of (a) averaged over one oscillation period from $B=-66$ mT to $-61.5$ mT. The colored dots correspond to the colored traces in (a).}
\end{figure}
Here we describe how we analyzed the data and extracted the phase information.
Figure \ref{fig:rawCoulomb} and Figure \ref{fig:rawKondo} show the raw data of the phase shift measurement in Fig. 1(c) and Fig. 3(a) and (b), respectively.
We display antiphase magneto oscillations of the two output currents and follow how the phase changes as a function of $V_{p}$.
However since each data set has a different nonoscillating background current, we extracted an oscillating component from the raw data  to emphasize the phase evolution as shown in Fig. 3(a).
First we determined a period of quantum oscillations by finding a peak in amplitude of a complex fast Fourier transform (FFT) of the raw data.
Secondly we obtained a nonoscillating background current by averaging the data over one oscillation period around each data point.
Finally we subtracted the background from the raw data and extracted the oscillating component.
This analysis allows us to clearly determine the phase evolution [Fig. 3(a)].

In addition, we performed a complex FFT of the oscillating component to obtain the phase information of the oscillation frequency as a numerical value [Fig. 1(c), Fig. 3(b), 3(c), and 3(d)].

\begin{figure}[htbp]
	\includegraphics[width=0.45\textwidth]{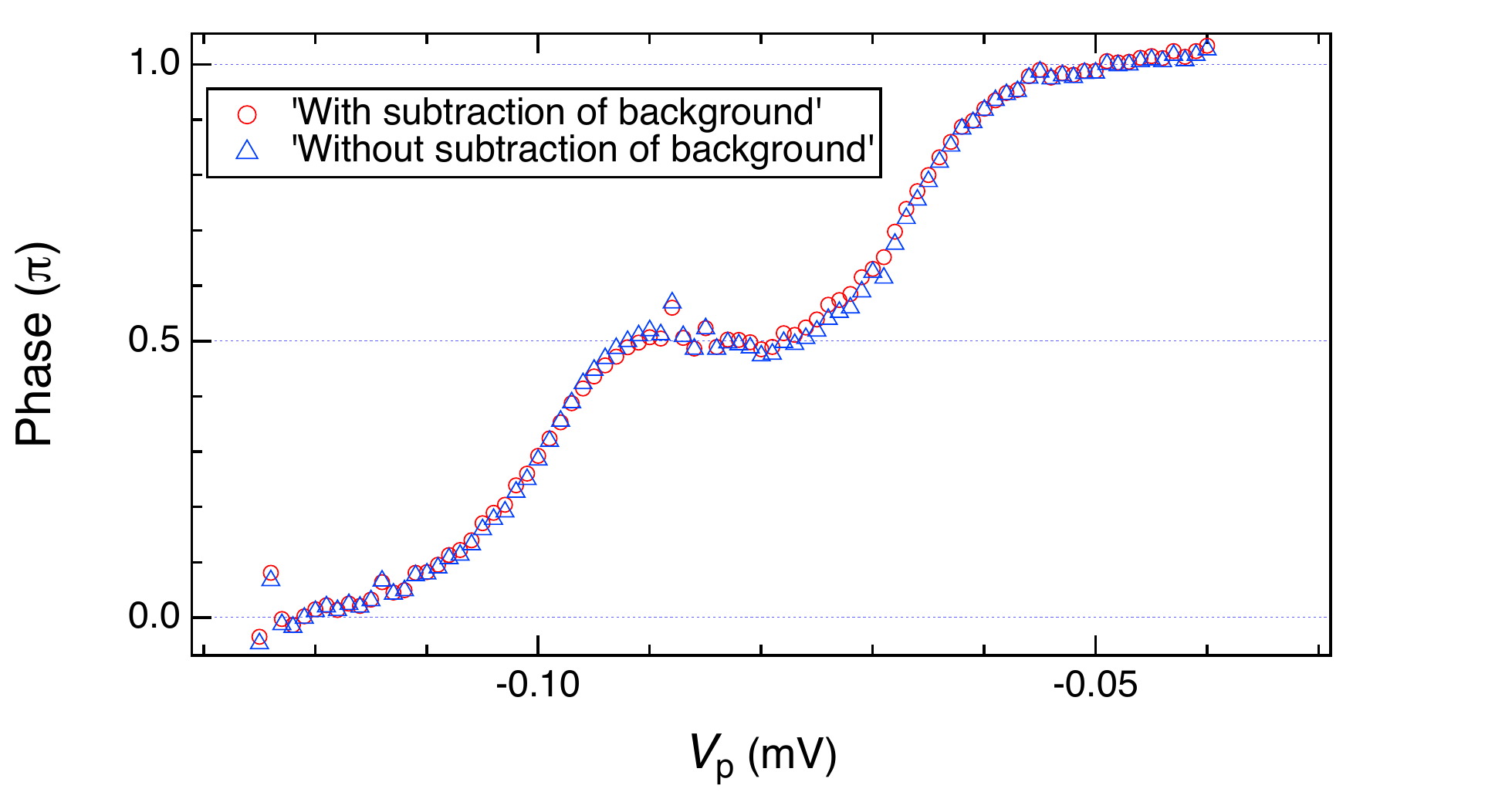}
	\caption{\label{fig:error}Comparison of the analyses with and without subtraction of the smoothed background currents. The data shown here is the same as the one in Fig. 3(c), which is used to evaluate the Kondo temperature. The red circles correspond the phase obtained with the subtraction and the blue triangles correspond the phase obtained without the subtraction.}
\end{figure}

It is also possible to obtain the phase information as a numerical value by performing a complex FFT of the raw data without subtraction of the background.
If there is a significant amplitude modulation while sweeping the magnetic field, the background subtraction may induce an extra error.
We have carefully checked that this effect has a negligible effect on the data analysis.
Figure \ref{fig:error} shows the comparison of the analyzed phase with and without subtraction of the background of the data shown in Fig. 3(c).
We observe only a very small difference between the two analyses [blue / red circles of Fig. \ref{fig:error}] due to fact that the amplitude of quantum oscillations is not constant over the swept magnetic field range and the background is nonlinear.
Such difference is considered to be an error of the analysis.
In our experiment this error is rather small as shown in Fig. \ref{fig:error} and only imposes an additional error bar of $\pm 10$ mK to the obtained Kondo temperature and electron temperature.

\section{\label{sec:s3}NRG calculation}

We performed NRG calculations \cite{Wilson1975b,Bulla2008b,Weichselbaum2007b,Weichselbaum2012b}
using the generalized Anderson Hamiltonian \cite[Eq.~(1)]{Hecht2009b}
\begin{multline}
  H = \smashoperator{\sum_{j,\sigma}} \varepsilon_{j} n_{j,\sigma} +
  \frac{U}{2} \smashoperator{\sum_{j,\sigma \neq j',\sigma'}}
  n_{j,\sigma} n_{j',\sigma'}
  \\
  + \smashoperator{\sum_{\alpha,k,\sigma}} \varepsilon_k
  c^\dagger_{\alpha,k,\sigma} c_{\alpha,k,\sigma} +
  \smashoperator{\sum_{j,\alpha,k,\sigma}} (t^j_\alpha
  c^\dagger_{\alpha,k,\sigma} d_{j,\sigma} + \text{H.c.})
\end{multline}
,where $j=1,2$ labels the dot levels,
$\sigma=\uparrow,\downarrow$ describes the electron spin,
$\alpha=\text{L},\text{R}$ denotes the channel index,
$\varepsilon_{1} < \varepsilon_{2}$ are level positions,
$U$ is the Coulomb interaction strength,
$t^j_\alpha$ are hopping amplitudes between level $j$ and channel $\alpha$,
and $\varepsilon_k$ denotes the dispersion relation of a flat band.
To avoid a proliferation of parameters, we choose the magnitudes of the tunneling matrix elements to be left-right symmetric and level independent, $|t^1_L| = |t^1_R| = |t^2_L| = |t^2_R|$, with relative sign $s=\sgn(t^1_{L} t^1_{R} t^2_{L} t^2_{R})$, and characterize the overall coupling strength in terms of the parameter $\Gamma = (\pi \rho)/2 \sum_{j,\alpha} |t^j_\alpha|^2$.
As the magnetic field is smaller than all other energy scales, we have neglected its presence.
Therefore, our calculations have fully exploited $\text{U}(1)$ charge and $\text{SU}(2)$ spin symmetries.  The discretization parameter was set to $\Lambda=4$, keeping approximately 6000 multiplets [corresponding to 20000 states] per iteration, which is required for the spinful two-channel calculation under consideration.
NRG produces the correlation function of any pair of local operators; using Kramers-Kronig relations yields the full Green's function.
Transmission amplitude and phase have been obtained from the complex quantity \cite[Eq.~(3)]{Hecht2009b}
\begin{equation}
t_{d\sigma}(T) = \int \mathrm{d}E \left(-\frac{\partial f_0(E,T)}{\partial E}\right)
\sum_{j,j',\sigma} 2\pi\rho t^j_L t^{j'}_R \mathcal{G}^R_{j,\sigma;j',\sigma}(E,T).
\end{equation}
For  Figs.~3(e)-(g) in the main paper, where $s=+$, the conductance through the dot is calculated using $\sum_\sigma {\rm Im}[t_{d \sigma}(T)]$, which equals the prediction of the Meir-Wingreen formula \cite{Meir1992b}.

\section{\label{sec:s4}Comparison between NRG calculations and experimental data}
\begin{figure*}[htbp]
	\includegraphics[width=0.9\textwidth]{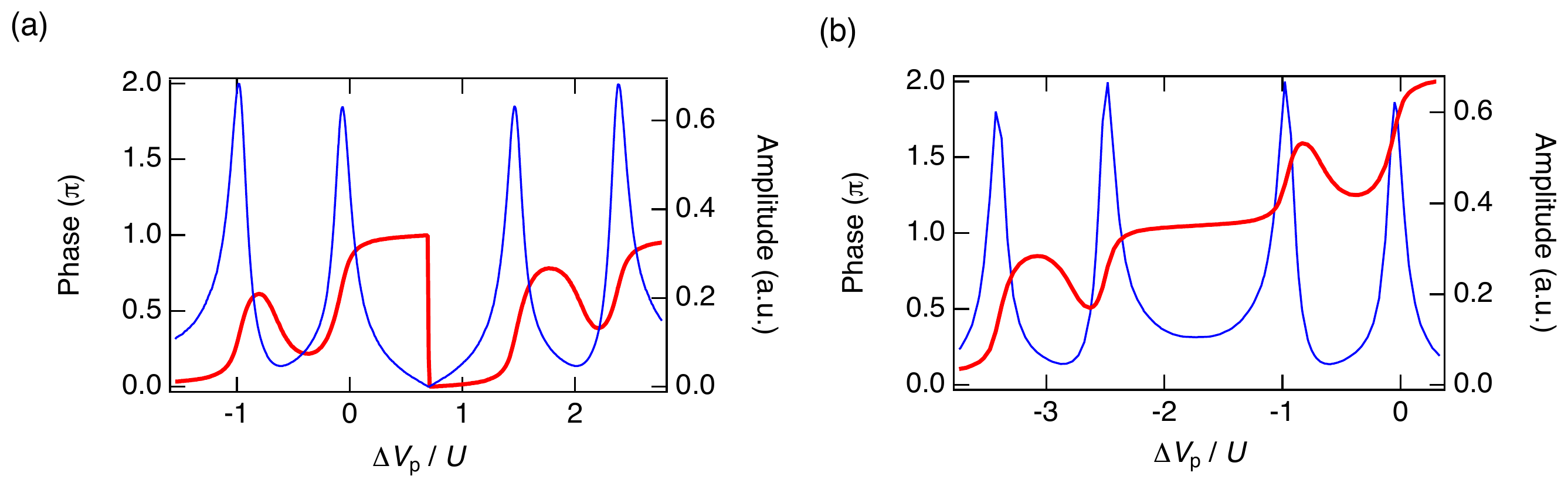}
	\caption{\label{fig:comp} Transmission phase (red) and transmission amplitude (blue) of a quantum dot calculated by a two-level Anderson model in the weak Kondo regime ($T\gg T_K$) for the case of $s=+$ (a) and $s=-$ (b). The fitting parameters, $\Gamma/U = 0.038$, $T/U=0.0055$ and $\delta / U = 0.5$ were used for both (a) and (b).}
\end{figure*}
In our fitting procedure using the numerical calculations of a two-level Anderson model, fitting parameters are the tunnel-coupling coefficients $t^j_{\alpha}$ between the dot level $j$ and the channel $\alpha$, the single level spacing $\delta$, the temperature $T$ and the bandwidth of the leads $D$, which are all given in units of the charging energy $U$.
In the whole fitting process $D$ is fixed to $20$ times bigger than $U$. Here the choice of $D$ does not change the result when it is large enough compared to all the other parameters.
To reduce the parameter space, we set $|t^1_L|=|t^1_R|=|t^2_L|=|t^2_R|$.
Prior to the fitting, the charging energy $U$ is first estimated. We obtained $U\sim 650 \ {\rm ~\mu eV}$ from the gate capacitance extracted from the Coulomb diamond at the base temperature and the Coulomb peak (CP) spacing at a temperature sufficiently high that the Kondo effect is so weak that the renormalization of the charging energy is negligible.
Another important parameter to be fixed before fitting is the relative sign of the tunnel-coupling coefficient for neighboring energy levels, $s_{j}= {\rm sgn}(t^{j}_{L}t^{j}_{R}t^{j+1}_{L}t^{j+1}_{R}$).
Figure \ref{fig:comp} shows transmission phase and amplitude calculated numerically by a two-level Anderson model at $T \gg T_K$ for different signs of $s$.
For $s=+$, a singular point called \textit{transmission zero} \cite{PhysRevLett.82.2358b} appears, which leads to a $\pi$-phase lapse and zero transmission amplitude when the Fermi level lies symmetrically between the two different levels [Fig. \ref{fig:comp}(a)].
On the other hand for $s=-$, there is no such singular point and the phase smoothly evolves between two different levels [Fig. \ref{fig:comp}(b)].
We assume from the current around the CPs shown in Fig. 2(a) of the main paper the relative sign between the Kondo level (governing the Kondo valley in the conductance) and the next-higher (next-lower) levels at more positive (negative) $V_{p}$ is $s_{j} = + \ (s_{j} = -)$. 
For our two-level Anderson model calculations, we therefore focus on the phase evolution of either a lower Kondo level with $s=+$, or an upper Kondo level with $s=-$ for comparison with the experimental data.
This assumption is consistent with the asymmetry of the phase evolution in the weak Kondo regime of the experiment [Fig. 3(d) in the main paper].

Here we describe the fitting procedure for $s=+$, whose results are presented in the main paper.
To determine the parameters used in Fig. 3(f), we first fitted the data of Fig. 3(c), where we can safely assume that the electron temperature is comparable to the fridge temperature of 110 mK, i.e. $T_e=110 {\rm \ mK}$.
We then tune only fitting parameters, $\Gamma$ and $\delta$.
To capture the most important feature of the phase evolution, we compared experimental data and calculations for the phase slope around the middle of the Kondo valley, $\Delta V_{p}\sim -20 {\rm ~mV}$.
To fit the experimental data, $\delta$ has to be large enough compared to $\Gamma$ (i.e. $\delta / \Gamma \gg 1$) and we confirmed numerically that the precise value of $\delta$ only plays a minor role for the phase behavior, as long as $\delta>0.4\ U$.
We therefore fixed $\delta$ to $0.5\ U$.
Note that experimentally $\delta$ can be evaluated via the excitation spectra.
However when $\Gamma$ is large enough to observe the Kondo effect, it is hard to find excited states and a precise evaluation of $\delta$ is not possible.
By varying $\Gamma$ as a single fitting parameter, we obtained $\Gamma \sim 62.4 {\rm ~\mu eV}$.
With these parameters, $T_K$ is determined to be approximately $50$ mK \cite{defTk}.
Let us emphasize that a different choice of $\delta$ does not lead to a significant change in $T_K$, instead, it leads to a slightly different value of $\Gamma$.
In such a way a very similar $T_K$ value is obtained.
The phase evolution at this temperature therefore allows for a precise evaluation of $T_K$, with an uncertainty of $\pm 10 \ {\rm mK}$.

To determine the parameters used in Fig. 3(e), we proceeded as follows: with $\Gamma = 62.4 {\rm ~\mu eV}$ and $\delta =  325{\rm ~\mu eV}$, temperature is then lowered as a fitting parameter to fit the phase curve in Fig. 3(b).
We note that for these $\Gamma/U$ values the $\pi/2$ plateau has still a finite and temperature dependent slope, even at zero temperature.
This allows for a relatively precise determination of the electron temperature of about 40 mK, somewhat higher than the fridge base temperature, but in good agreement with former measurements on 2DEGs in the same electromagnetic environment \cite{PhysRevB.81.245306b}.

Finally, to determine the parameters used in Fig. 3(g), we proceeded as follows: with $T=40 {\rm \ mK}$ and $\delta =  325{\rm ~\mu eV}$, $\Gamma$ is varied to fit the data in Fig. 3(d).
The data is fit best for $\Gamma = 24.7{\rm ~\mu eV}$.
The determination of the exact Kondo temperature in this weak Kondo regime is more delicate as the exponential dependence on $\Gamma/U$ leads to relatively large uncertainties.
Nevertheless we estimate $T_K$ to be more than 30 times smaller than the actual electron temperature. Finally let us mention that the values for $\Gamma$ extracted from the NRG fitting are much smaller than if one would take simply the width of the CPs (in our case $24.7 {\rm ~\mu eV}$ compared to $130 {\rm ~\mu eV}$).
The asymmetry in the phase evolution in Fig. 3(g) around $\Delta V_{p} / U = -0.5$ originates from the influence of the nearby level and reflects the orbital parity, $s_{j}=+$.
We also performed the same fitting procedure with the other choice of sign, $s_{j}=-$, which led to very similar parameters.

For both fitting procedures we get remarkable agreement between the experimental data and the NRG calculations for the phase evolution.
The $\pi/2$-phase shift persisting up to $T_K$ [Fig. 3(b)] clearly indicates that coherent current flows through the Kondo resonance once a localized spin is formed in the quantum dot for $T\leq T_K$.
Compared to the phase, agreement between the current $I_{\rm 2}$ in the experiment and the conductance $G$ in the calculations is not as good.
This may arise from several reasons: one possibility is that in our calculation we did not include left-right asymmetry of the tunnel coupling between the leads and the quantum dot whereas for the actual quantum dot this asymmetry exists.
Another possible reason is that in the experiment there is a linear background from the upper path of the interferometer whereas the calculated conductance $G$ does not contain such a contribution.

In the gate voltage region ($V_{p}$) in which the current $I_{\rm 2}$ shows the Coulomb blockade peak on the right, the phase increases towards one with increasing $V_{p}$ occurs somewhat sooner (relative to the position of the CP) for the experimental data in Fig. 3(b) and 3(c) than for the NRG calculations in Fig. 3(e) and 3(f).
This deviation between experiment and theory in the tails of the phase curves may be due to the fact that we experimentally extract the phase from the \textit{coherent} part of the current whereas the \textit{total} current $I_{\rm 2}$ that is plotted in Figs. 3(b) - 3(d) also contains the incoherent part. In this context, further theoretical studies are necessary.

\section{\label{sec:s5} Comparison to previous measurements of the transmission phase shift in the Kondo regime}
As briefly pointed out in the introductory paragraph, earlier measurements have attempted to observe the theoretically predicted transmission phase shift of $\pi/2$ in the Kondo regime when scanning through two successive CPs. 
In the pioneering experiment of Ji et al. \cite{Ji2000b}, however, \textit{``the total accumulated phase as the two spin-degenerate levels cross the Fermi energy in the leads is nearly 2$\pi$ with a phase shift $\pi$ in the Kondo valley"} in contradiction with Kondo theory.
In a followup  experiment \cite{Ji2002b}, where the quantum dot was tuned into the unitary limit, an \textit{``even more peculiar and unexpected behavior"} was found.
\textit{``The transmission phase was observed to evolve almost linearly over a range of about $1.5\pi$ when the Fermi energy was scanned through a spin degenerate energy level of the QD"} and from these results it was suggested \textit{``that a full explanation of the Kondo effect may go beyond the framework of the Anderson model."}.
  
In a more recent experiment, the same group \cite{Zaffalon2008b} claim to have observed the theoretically predicted $\pi/2$-phase shift in the Kondo regime \cite{Gerland2000b}.
However, in our opinion the published data do not support this claim, as we now explain.
The puzzling feature of this experiment is the fact, that the transmission phase shift of $\pi/2$ is observed in the Kondo valley for temperatures much larger than the Kondo temperature and that this $\pi/2$ transmission phase shift persists over the entire Kondo valley, that is when moving away from the center of the Kondo valley towards the CPs, the $\pi/2$ phase shift is unchanged as seen in Fig. 3 of ref. \cite{Zaffalon2008b}.
Both of these observations are in contradiction with theory \cite{Gerland2000b}. 

Indeed, Kondo theory predicts that the total phase shift across two CPs is $\pi$ and the appearance of a $\pi/2$ plateau around the center of the Kondo valley for $T < T_K$.
This $\pi/2$ plateau is a property of the strong-coupling fixed point and the accompanying Fermi-liquid regime of the Kondo model, i.e. of the regime $T / T_K < 1$ \cite{Nozieres:xyb}.
Strictly speaking, the Fermi-liquid regime is limited to $T / T_K \ll 1$, but the $\pi/2$ plateau was found to persist to about $T / T_K = 1$ \cite{Gerland2000b, Silvestrov2003b, Hecht2009b}.
On the other hand, the plateau will disappear for $T / T_K \gg 1$ as expected within Kondo theory and demonstrated in the present work. For temperatures above the Kondo temperature, the phase shift is enhanced on one side of the center of the Coulomb valley while it is decreased on the other side of the Coulomb valley.

Such a feature is indeed observed in Fig. 4 of ref \cite{Zaffalon2008b}.
When changing $\Gamma$ (which is equivalent to changing the Kondo temperature), the phase shift very close to the left CP starts to climb well above $\pi/2$ while it starts to fall well below $\pi/2$ close to the right CP.
In the center between the two CPs no data is shown due to too small signal to noise ratio.
Carefully inspecting Fig. 3 of ref. \cite{Zaffalon2008b}, one actually sees a very similar tendency:  the phase very close to the left CP climbs above $\pi/2$ just before the conductance drops close to zero.
A similar, but opposite tendency is seen close to the right CP where the phase drops below $\pi/2$.
On the other hand, the almost unchanged $\pi/2$ plateau in the entire region between the two CPs is very surprising as the data of Ref. \cite{Zaffalon2008b} does not show any Kondo signature near the center of the Kondo valley such as the zero bias anomaly or a logarithmic increase of the conductance when decreasing temperature.
Also the fact that the conductance is close to zero in the \textit{entire} Kondo valley makes a reliable evaluation of the phase impossible.
Comparing the phase data of Fig. 3 of ref. \cite{Zaffalon2008b} with Fig. 3d of our manuscript, one has the impression that the phase data outside the CPs, where still a good signal noise ratio is achieved due to sufficient conductance, is in line with our findings.
The phase climbs above $\pi/2$ close to the left CP while it falls below $\pi/2$ close to the right CP.
Such a behavior is indeed expected for  a temperature regime ($T \gg T_K$).
Only the unchanged $\pi/2$ plateau in the Kondo valley of Ref . \cite{Zaffalon2008b}  seems to be inconsistent with Kondo theory.

The difficulty of such transmission phase measurements lies in the fact that the contribution of multipath interferences is very hard to determine. Usually this is done by checking whether the phase shifts smoothly when moving one of the side gates of the interferometer. This indeed verifies whether the phase rigidity is lifted but does not tell whether multi path contribution are suppressed. For standard AB interferometers, this can only be checked from the harmonic content of the Fourier spectrum, which however is a very insensitive method. 
The important advantage of our interferometer is that we can directly evaluate the contribution of multiple paths. 
This can be done by verifying the antiphase oscillations of the two output currents.
Our experiment therefore constitutes a \textit{true} two-path interferometer.
The high quality data in the Kondo regime allows us to perform a detailed and quantitative analysis that covers a temperature range from $T / T_K < 1$ to $T / T_K \gg 1$ and is consistent with theory, providing a direct evidence of the $\pi/2$-phase shift and the screened singlet ground state.

%


\begin{thebibliography}{16}
\bibitem [{def()}]{defTk}%
  \BibitemOpen
  \href@noop {} {}\bibinfo {note} {The Kondo temperature is calculated by
  $T_{\rm K} = \chi_0 / 4$, where $\chi_0$ is the local spin susceptibility at
  zero temperature.}\BibitemShut {Stop}%
\end{thebibliography}
\end{document}